\documentclass[amsmath,amssymb,showkeys,superscriptaddress,aps,prd,10pt,onecolumn]{revtex4-2}
\usepackage{graphicx}
\usepackage[utf8]{inputenc}
\usepackage[caption=false]{subfig}
\usepackage{multirow}
\usepackage{appendix}
\usepackage{graphicx,epstopdf}

\usepackage{colortbl}
\usepackage{xcolor}
\usepackage[colorlinks=true, urlcolor=blue, linkcolor=blue, citecolor=blue]{hyperref}
\usepackage{float}
\usepackage{adjustbox}

\begin{document}

\title{Mass spectrum and magnetic moments of singly-charmed baryons: a quark-diquark model analysis of $\Omega_{c}(3185)^0$ and $\Omega_c(3327)^0$}

\author{Sinem Küçükyılmaz}%
\email[]{24280455@stu.omu.edu.tr }
\affiliation{Department of Physics, Faculty of Science, Ondokuz Mayis University, 55200, Samsun, Türkiye }

\author{Halil Mutuk}%
\email[]{hmutuk@omu.edu.tr }
\affiliation{Department of Physics, Faculty of Science, Ondokuz Mayis University, 55200, Samsun, Türkiye }
 
\begin{abstract}
Research on singly-heavy baryons, especially those with a charm quark, offers a distinct perspective on the non-perturbative behavior of Quantum Chromodynamics (QCD). In this work, we investigate the recently observed $\Omega_{c}(3185)^{0}$ and $\Omega_{c}(3327)^{0}$ as singly-charmed baryons within the framework of the quark-diquark model. By employing a non-relativistic method with a Cornell-like potential, we systematically determine magnetic moments and mass spectra. Our analysis reveals that the $\Omega_{c}(3185)^{0}$ can be effectively described as a $2S$ state with quantum numbers $J^{P}=\frac{1}{2}^{+}$ or $\frac{3}{2}^{+}$, or alternatively as a $1P$ state with $J^{P}=\frac{1}{2}^{-}$ or $\frac{3}{2}^{-}$, depending on the diquark configuration. Similarly, the $\Omega_{c}(3327)^{0}$ is consistent with a $2S$ configuration. We also investigate their magnetic moments, emphasizing the critical role of diquark correlations in shaping the electromagnetic properties of these states. Our results not only validate existing theoretical models but also offer new insights into the nature of singly-heavy baryons, setting the stage for future experimental and theoretical investigations in heavy baryon spectroscopy. This paper emphasizes the importance of diquark configurations in elucidating the mass spectrum and electromagnetic characteristics of singly-charmed baryons, aiding in the broader effort to decipher QCD intricacies.
\end{abstract}

\maketitle

\section{Motivation}
According to the quark model, baryons are a family of subatomic particles composed of three quarks \((qqq)\) or three antiquarks \((\bar{q} \bar{q} \bar{q})\). The differing mass scales of quarks make singly-heavy baryons \((Qqq)\) - composed of a heavy quark \((Q)\)and two light quarks \((q)\) - a compelling subject of study. Researchers have identified numerous singly-heavy baryons and determined the quantum numbers of most known states. Recent experimental advances in heavy baryon physics have enabled deeper insights into these states. The resonance energies probe quark interactions within baryons, much like atomic spectroscopy, where spectral lines reveal electron-nucleus dynamics.

Experimental of conventional heavy baryons provide valuable insights into the non-perturbative regime of QCD \cite{Brambilla:2019esw, Chen:2022asf}. A singly-heavy baryon \((Qqq)\) consists of a heavy quark \((Q)\) and two light quarks \((qq)\), bound by gluon-mediated interactions. This system serves as a QCD analog to the helium atom, where two light electrons orbit a nearly stationary nucleus, bound by electromagnetic forces. Due to their unique structure, heavy baryons have attracted significant theoretical and experimental interest in recent years \cite{ARGUS:1993vtm, E687:1993bax, Ammosov:1993pi, CLEO:1994oxm, CLEO:1995amh, CLEO:1996czm, CLEO:1996zcj, ARGUS:1997snv, E687:1998dwp, CLEO:1999msf, CLEO:2000mbh, CLEO:2000ibb, Belle:2004zjl, BaBar:2006itc, Belle:2006xni, Belle:2006edu, BaBar:2006pve, BaBar:2007zjt, BaBar:2007xtc, LHCb:2012kxf, LHCb:2017jym, LHCb:2017uwr, LHCb:2018haf, LHCb:2018vuc, LHCb:2019soc, LHCb:2020xpu, LHCb:2020tqd, LHCb:2020iby, LHCb:2020lzx, CMS:2021rvl}.

Recent results from the LHCb~\cite{LHCb:2023sxp} have established the existence of two novel singly-charmed states, $\Omega_{c}^{0}(3185)$ and $\Omega_{c}^{0}(3327)$, observed in the $\Xi_{c}^{+}K^{-}$ final state with the following characteristics:
\begin{eqnarray}
M_{\Omega_c(3185)} &=& 3185.1 \pm 1.7 ^{+7.4}_{-0.9} \pm 0.2 \, \text{MeV}, \nonumber \\
\Gamma_{\Omega_c(3185)} &=& 50 \pm 7 ^{+10}_{-20} \, \text{MeV}, \\
M_{\Omega_c(3327)} &=& 3327.1 \pm 1.2 ^{+0.1}_{-1.3} \pm 0.2 \, \text{MeV}, \nonumber \\
\Gamma_{\Omega_c(3327)} &=& 20 \pm 5 ^{+13}_{-1.0} \, \text{MeV}.
\end{eqnarray}

Although masses and widths of these two excited states have been measured with unprecedented precision, their spin quantum numbers remain undetermined. The quark content of these states is \(css\), classifying them as singly-charmed baryons. Owing to the significant mass difference between the charm quark and the strange quarks, the mass spectrum of singly-charmed baryons can be accurately described using heavy quark effective theory (HQET) \cite{Grozin:1992yq}. Investigations of singly-charmed baryon spectra yield significant insights into the nature of strong interactions and hadronic spectroscopy, offering profound implications for our understanding of the fundamental principles of QCD.

The observation of these two new excited states has prompted numerous theoretical investigations, interpreting them as both conventional baryons \cite{Yu:2023bxn, Luo:2023sra, Wang:2023wii, Karliner:2023okv, Ortiz-Pacheco:2023kjn, Pan:2023hwt} and molecular pentaquark configurations \cite{Feng:2023ixl, Yan:2023ttx, Xin:2023gkf, Yan:2023tvl, Ozdem:2023okg}. In Ref. \cite{Yu:2023bxn} the \(^{3}P_{0}\) model was employed to systematically analyze singly-heavy baryons, including \(\Lambda_{Q}\), \(\Sigma_{Q}\) and \(\Omega_{Q}\). The analysis supports identifying the $\Omega_{c}(3185)^0$ as a radial excitation ($2S$, $J^P=3/2^+$) and the $\Omega_{c}(3327)^0$ as an orbital excitation ($1D$, $J^P=3/2^+$), in agreement with quark model expectations for charmed baryons. The \(\Omega_c(3327)^0\) has been interpreted as a $css$ \(1D\) wave state within the Gaussian expansion method in Ref. \cite{Luo:2023sra}. Meanwhile, Ref. \cite{Wang:2023wii} employed QCD sum rules to calculate its mass and pole residue, assigning it to a \(D\)-wave state with possible quantum numbers \(J^P = 1/2^+\), \(3/2^+\), or \(5/2^+\). In contrast, Ref. \cite{Karliner:2023okv} proposed the \(\Omega_{c}(3185)^0\) and \(\Omega_c(3327)^0\) as \(2S_{1/2}\) and \(2S_{3/2}\) states, respectively. Their analysis revealed that while the $1S-2S$ splitting is smaller than expected and the hyperfine splitting is larger, these findings do not invalidate the proposed assignments. Additionally, Ref. \cite{Ortiz-Pacheco:2023kjn}  presented comprehensive results on the mass spectra and electromagnetic couplings for $S$ and $P$ - wave states across quark-like ${(Qqq)}$, double-heavy ${(Qqq)}$, and triple-heavy ${(QQQ)}$ baryon sectors. The findings suggest that the $\Omega_{c}(3185)^0$ state appears to be for the \(\rho\)-excited state of the $P$-wave $(J^P=3/2^-)$. The mass spectra of the baryons \((Q = c, b)\) \(\Sigma_{Q}\), \(\Xi^{\prime}_{Q}\), and \(\Omega_{Q}\) are investigated using the linear Regge trajectory and perturbative treatment method inside a quark-diquark configuration \cite{Pan:2023hwt}. By considering states with $J^{P}=1/2^{+}$ and $J^{P}=3/2^{+}$, the analysis identifies the $\Omega_{c}(3185)^{0}$ as a $2S$ state and the $\Omega_{c}(3327)^{0}$ as a $1D$ state.

Assuming the \(\Omega_{c}(3185)^0\) and \(\Omega_c(3327)^0\) states to be molecular structures composed of \(S\)-wave \(D \Xi\) and \(D^* \Xi\) respectively, Ref. \cite{Feng:2023ixl} investigated their strong decay patterns. Their analysis supports identifying $\Omega_c(3327)^0$ as a $J^P=3/2^-$ hadronic molecule in the $D^*\Xi$ channel, while \(\Omega_{c}(3185)^0\) is interpreted as a molecule that has meson-baryon a significant \(D \Xi\) component. Ref. \cite{Yan:2023ttx} employed a phenomenological contact-range interaction model to describe these charmed meson molecular states, assigning \(\Omega_{c}(3185)^0\) and \(\Omega_c(3327)^0\) as $D\Xi$ and $D\Xi$ molecular states with $J^P = 1/2^-$ and $3/2^-$, respectively. Meanwhile, Ref. \cite{Xin:2023gkf} conducted a comprehensive QCD sum rules analysis of these states mass spectra and pole residues, treating them as bound molecular pentaquark configurations. 
Using the quark delocalization color screening model, Ref. \cite{Yan:2023tvl} investigated excited \(\Omega_c\) states and suggested the \(\Omega_{c}(3185)^0\)  could be interpreted as a \(D \Xi\) molecular state with \(J^P = 1/2^-\). In a complementary approach, Ref. \cite{Ozdem:2023okg} employed light-cone QCD sum rules to study the magnetic moments of these states. Their analysis characterized the $\Omega_{c}(3185)^0$ as an $S$-wave $D \Xi$ hadronic molecule and the $\Omega_c(3327)^0$ as an $S$-wave  $D^* \Xi$ molecular pentaquark configuration.

In principle, diquark \(qq\) correlations can significantly influence the hadronization process, giving rise to the concept of compact tetraquarks \cite{Lee:2007tn, Maiani:2004vq}. Experimental measurements of light-quark pair correlation densities further reveal strong interactions within so-called ``good" diquark configurations \cite{Alexandrou:2006cq}.  In this context, charmed baryons containing two light quarks \(Qqq\) represent an ideal testing ground for probing these diquark dynamics.

The literature reveals divergent interpretations: while some studies support conventional baryon descriptions, others favor exotic configurations. Although existing predictions agree with experimental measurements within uncertainties, the varied theoretical approaches underscore the need for further investigation to resolve these states internal structure. In this work, we employ the quark-diquark model to systematically calculate the masses and magnetic moments of the \(\Omega_{c}(3185)^0\) and \(\Omega_c(3327)^0\), aiming to shed light on their underlying quark organization.

This paper is organized as follows: Section \ref{model} introduces the theoretical framework of our study. In Section \ref{numeric}, we present our numerical results, including the calculated masses and magnetic moments. Finally, Section \ref{conclusion} summarizes our key findings and discusses their implications for understanding the nature of these baryonic states.

\section{Quark-Diquark Model}\label{model}

The diquark concept was  originally proposed as a complementary approach to understanding baryon structure \cite{Ida:1966ev, Lichtenberg:1967zz, Lichtenberg:1969sxc}. In this framework, baryons are modeled as bound states of a quark and a diquark, where diquarks emerge as correlated quark pairs $qq$ in non-singlet color configurations. The quark-diquark model has proven particularly effective for calculating baryon spectra and form factors, as demonstrated in numerous theoretical studies  \cite{Santopinto:2004hw, Ferretti:2011zz, Santopinto:2014opa, DeSanctis:2014ria}. For a thorough review of diquark properties and applications, see \cite{Barabanov:2020jvn} and references therein.

From a group-theoretical perspective, the quark-quark interaction exhibits an attractive channel in the color \(\bar{3}\) representation of SU(3). This attraction leads to diquark formation, where the resulting diquark possesses color properties analogous to an antiquark, reflecting a fundamental quark-antiquark symmetry. Within the baryon structure, this diquark appears to the remaining quark as an effective color source equivalent to an antiquark. In doubly heavy baryons \((QQq)\), the heavy quark pair \((QQ)\)  preferentially forms a compact diquark. The experimental observation of $\Xi_{cc}^{++}$ ($ccu$) with mass $m_{\Xi_{cc}^{++}} = 3621.40 \pm 0.72$ MeV \cite{LHCb:2017iph, LHCb:2018pcs} provides compelling evidence for such \(cc\) diquark formation. Conversely, in singly-heavy baryons \((Qqq)\), the light quark pair \((qq)\) is more likely to form the diquark component. The quark-diquark model has been shown to be particularly effective for describing baryon excitation spectra, as demonstrated in studies of light-quark baryons \cite{Richard:1992uk}. Non-relativistic quark model analyses \cite{Fleck:1988vm} reveal that diquark clustering occurs prominently in both \((qqq)\) and \((Qqq)\) baryons with high orbital angular momentum. These studies further establish that when \(m_Q / m_q \gg 1\), a \((qq)\) diquark dominates, while for \(m_Q / m_q \simeq 1\), a \((Qq)\) diquark becomes more favorable a scenario we also consider in our current investigation.

While the quark-diquark model has been extensively employed and has yielded valuable insights into baryon spectroscopy \cite{Ida:1966ev, Lichtenberg:1967zz, Lichtenberg:1969sxc, Santopinto:2004hw, Ferretti:2011zz, Santopinto:2014opa, DeSanctis:2014ria,Barabanov:2020jvn}, its validity remains a subject of ongoing debate. Recent studies have challenged the model's universality, questioning whether the diquark approximation fully captures baryon structure \cite{Gross:2022hyw, Zhong:2024mnt, Li:2025frt}. These works highlight the importance of examining more complex quark dynamics beyond two-body correlations, advocating for either complete three-quark treatments or alternative configurations like molecular models. Despite these challenges, the quark-diquark approach remains a computationally tractable and phenomenologically successful framework, particularly for heavy baryons where the mass hierarchy enhances diquark clustering effects. In this context, our work systematically evaluates the model consistency with the recently observed $\Omega_c^0 (3185)$ and $\Omega_c^0 (3327)$ states through simultaneous analysis of their mass spectra and magnetic moments, thereby contributing to this critical discourse.

We investigate the \(\Omega_{c}(3185)^0\) and \(\Omega_c(3327)^0\) states within a non-relativistic quark-diquark framework. In this approach, baryons are modeled as bound states of a constituent quark and a compact diquark, where the diquark is treated as a pointlike object. For the quark-diquark interaction, we adopt the Cornell potential, which has the form:

\begin{equation}
V(r)=V_V(r)+V_S(r)=\frac{\kappa \alpha_s}{r} + br, \label{Cornell}
\end{equation}
In the potential model formalism, the color factor is denoted by \(\kappa\), while \(\alpha_s\) represents the strong coupling constant, often referred to as the QCD analog of the electromagnetic fine-structure constant. The string tension, characterizing the linear confinement behavior, is denoted by \(b\). As expressed in Eq.(\ref{Cornell}), the potential \(V_V(r)\) corresponds to the vector part of the interaction, originating from one-gluon exchange (OGE) and possessing a Lorentz vector structure. In contrast, the confining potential \(V_S(r)\) arises from long-range interactions and is assumed to have a Lorentz scalar structure.

The Schrödinger equation can be separated into angular and radial components using spherical coordinates in the center of mass frame. The reduced mass parameter $\mu$ is defined as
\begin{equation}
\mu \equiv \frac{m_1 m_2}{m_1 + m_2},
\end{equation}
where the effective masses of the constituent quark components in this binary system are represented by the parameters $m_1$ and $m_2$. The radial component of the Schrödinger equation takes the following form:
\begin{equation}
    \left\{  \frac{1}{2\mu}\left[-\frac{\text{d}^2}{\text{d}r^2}+\frac{\ell(\ell+1)}{r^2}\right] +V(r) \right\} \psi(r) = E\psi(r).
    \label{schtrans}
\end{equation}
Using the Breit-Fermi Hamiltonian for OGE \cite{Lucha:1991vn}, the spin-spin interaction term may be included in the potential
\begin{align}
    V_{SS}(r) &= -\frac{2}{3(2\mu)^2}\nabla^2 V_V\langle \mathbf{S}_1\cdot\mathbf{S}_2 \rangle \nonumber\\
    &= -\frac{2\pi\kappa\alpha_S}{3\mu^2}\delta^3(r) \langle\mathbf{S}_1\cdot\mathbf{S}_2 \rangle.
    \label{spinspin}
\end{align}

A Gaussian function, which serves as a smearing function, can be used to define the spin-spin interaction in the zeroth-order (unperturbed) potential in place of the Dirac delta function:
\begin{equation}
    V_{SS}(r) = -\frac{2\pi\kappa\alpha_S}{3\mu^2}\left(\frac{\sigma}{\sqrt{\pi}}\right)^3\exp\left(-\sigma^2 r^2\right) \langle \mathbf{S}_1\cdot\mathbf{S}_2 \rangle,
    \label{spinsmear}
\end{equation}
where this introduces a new parameter \(\sigma\). The modified potential is given by the following expression:

\begin{equation}
V_{\text{eff}}(r)=V_V(r)+V_S(r)+V_{SS}(r). \label{effpot}
\end{equation}

The Schrödinger equation can now be expressed as follows:
\begin{equation}
    \left[  -\frac{\text{d}^2}{\text{d}r^2} +\tilde{V}_{\text{eff}}(r)\right]\varphi(r) = 2\mu E\varphi(r).
    \label{schtranss}
\end{equation}
Here effective interparticle potential $\tilde{V}_{\text{eff}}(r)$ takes the following form:
\begin{equation}
\tilde{V}_{\text{eff}}(r)\equiv 2\mu\left[ V_V(r)+V_S(r)+V_{SS}(r) \right] + \frac{\ell(\ell+1)}{r^2}.
    \label{veff}
\end{equation}

The total interaction potential consists of three physically different components: the Cornell potential $(V(r))$, spin-spin interactions $(V_{SS}(r))$, and orbital terms $(V_{\text{orb}})$.


The quark-diquark model is a theoretical framework that facilitates the understanding of the three-body problem for singly-heavy baryons. By carefully accounting for quark masses, this model reduces the problem to a two-step process. In the second step, the mass of the quark-diquark bound state is determined. It is assumed that all excitations occur between the diquark and the quark. There are no internal excitations of the quark itself, as the excitations in the quark-diquark model used here are confined to the quark-diquark bound system. The following mass spectrum is obtained by solving the non-relativistic Schrödinger equation with the above effective potential:

\begin{equation}
M_B=m_{d} + m_q + E.
\end{equation}
Here, \(m_{d}\) represents the mass of the diquark, \(m_q\) is the mass of the quark, and \(E\) denotes the energy eigenvalue of the quark-diquark system.

As is well known, hadrons exist only in color singlet states, meaning they must be colorless (color neutral). The fundamental difference between $q\bar{q}$ and $qq$ systems arises from their distinct color representations. In \(SU(3)\) symmetry, a quark and an antiquark combine according to the decomposition \(\vert q \bar{q} \rangle \to 3 \otimes \bar{3} = 1 \oplus 8\), where the color factor for this representation is \(\kappa = -4/3\). Under SU(3)$_c$ color symmetry, the two-quark system decomposes as:
\begin{equation}
\mathbf{3} \otimes \mathbf{3} = \overline{\mathbf{3}} \oplus \mathbf{6},
\end{equation}
where the symmetric color sextet configuration is indicated by $\mathbf{6}$ and the antisymmetric color antitriplet state is represented by $\overline{\mathbf{3}}$. The color factor is \(\kappa = -2/3\) for the antitriplet state, which is attractive, and \(\kappa = +1/3\) for the sextet state, which is repulsive. Consequently, we only consider diquarks in the antitriplet state. 

It is crucial to acknowledge that modifying the color factor from \(\kappa = -4/3\) (for quark-antiquark in the color singlet state) to \(\kappa = -2/3\) (for quark-quark in the antitriplet color state) is equivalent to introducing a factor of \(1/2\) in the Coulomb part $(V_V(r))$ of the potential. This factor is often treated as globally because it is derived via the wave-function color structure \cite{Ebert:2007rn, Lu:2016zhe, Mutuk:2021epz}. Thus, \(b \to b/2\) and \(\kappa \to \kappa/2\) are the general rules for getting the energy eigenvalues of diquark systems.

\section{Numerical Results and Discussion}\label{numeric}
The model incorporates five parameters: the constituent charm quark mass \(m_c\), the constituent strange quark mass \(m_s\), the coupling constant \(\alpha_s\), the string tension \(b\), and the smearing parameter \(\sigma\). We fitted these parameters to experimental data for charmed-strange mesons from the PDG \cite{Workman:2022ynf}, excluding states not listed in the summary tables. The \(D_{s0}^\ast(2317)\) state was also omitted from the fit due to its inconsistency with quark-model predictions.

We minimize the function
\begin{equation}
\chi^2 = \sum_{i=1}^{N_{\rm data}} w_i{\left[M_{\text{exp},i}-M_{\text{model},i}\right]^2}.
\label{eq:chisqu}
\end{equation}
Here, \(M_{\text{exp}}\) and \(M_{\text{model}}\) denote the experimental and theoretical masses, respectively, and \(w_i\) is the weight function for each state. We adopt \(w_i = 1 \, \text{MeV}^{-1}\), giving equal statistical weight to all input states. The fitted model parameters are listed in Table \ref{tab:ParametersFit}. Using these parameters, we compute the charmed-strange meson mass spectra, shown in Table \ref{tab:mesonPDG}. As clear from the table, the calculated meson masses agree well with the experimental data. 

\begin{table}[H]
\caption{The quark model parameters determined by ﬁtting the charmed-strange meson mass spectra.}
    \label{tab:ParametersFit}
    \centering
    \begin{tabular}{cccccc}
    \hline
    \hline 
   $m_c~[{\rm GeV}]$&$m_u~[{\rm GeV}]$& $\alpha_s$ & $b~[{\rm GeV}^2]$ & $\sigma~[{\rm GeV}]$ & $\chi^2$\\
        \hline
         1.43383 & 0.33071 & 0.68353 & 0.11139 & 0.45608 &0.002924\\
         \hline
        \hline
    \end{tabular}
\end{table}

\begin{table}[H]
\caption{Charmed-strange meson mass spectra from the parameters obtained in this work.  The mass results are in unit of MeV.}
    \label{tab:mesonPDG}
    \centering
    \begin{tabular}{ccc}
    \hline
    \hline   Meson  &This work & Experiment  \\
        \hline 
       $D_s^{\pm}$  & 1986 &$1968.35 \pm 0.07$ \\
       $D_s^{* \pm}$ & 2113 & $2112.2 \pm 0.4$\\
       $D_{s1}^{\pm}(2536)$ & 2509 & $2535.11 \pm 0.06$ \\
       $D_{s2}^{*}(2573)$ & 2534 & $2569.1 \pm 0.8$\\
       $D_{s1}^{*\pm}(2700)$& 2719 & $2714 \pm 5$ \\
        \hline
        \hline
    \end{tabular}
\end{table}

\subsection{Mass Spectrum}

The quark content of the \(\Omega_{c}(3185)^0\) and \(\Omega_c(3327)^0\) states is expected to be \(css\). In the quark-diquark model, this configuration can arise in various ways. For \(\Omega_c\) states, diquark formation may occur as
\begin{eqnarray}
\Omega_c^1 &=& \left[cs \right]s, \nonumber \\
\Omega_c^2 &=& \left\lbrace cs \right\rbrace s, \nonumber \\
\Omega_c^3 &=& \left\lbrace ss \right\rbrace c,
\end{eqnarray}
where \(\left[ ab \right]\) represents a \(J^P = 0^+\) diquark that is antisymmetric under the exchange \(a \leftrightarrow b\), while \(\left\lbrace ab \right\rbrace\) denotes a \(J^P = 1^+\) diquark that is symmetric under \(a \leftrightarrow b\) \cite{Yin:2019bxe}.

For the baryons studied here, we numerically solve Eq.(\ref{schtranss})first for the diquark system and then for the quark-diquark system. Tables~\ref{tab:swave} and~\ref{tab:pwave} present the calculated masses for \(S\)-wave and \(P\)-wave configurations, respectively. These results provide key insights into the mass spectra and magnetic moments of \(\Omega_{c}(3185)^{0}\) and \(\Omega_{c}(3327)^{0}\) within the quark-diquark model framework.

As shown in Table \ref{tab:swave}, our calculated mass for the \(1^2S_{1/2}\) state of \(\Omega_c^1\) agrees well with several theoretical approaches: (i) a non-relativistic quark model focusing on spin-dependent interactions \cite{Shah:2016mig}, (ii) a relativistic quark-diquark treatment of confinement effects \cite{Ebert:2011kk}, (iii) a hybrid HQET-quark model exploiting heavy quark symmetry \cite{Roberts:2007ni}, (iv) Regge trajectory analyses \cite{Valcarce:2008dr},and (v) a Gaussian expansion method (GEM) quark model \cite{Yoshida:2015tia}. While the \(1S\) of \(\Omega_c^1\) and \(\Omega_c^2\) show consistent agreement across these studies, \(\Omega_c^3\)exhibits notable deviations. This discrepancy likely stems from its unique quark structure, where the axial-vector \(\{ss\}\) diquark couples with the charm quark.

Comparative analysis with existing theoretical studies shows notable variations in the predicted masses of orbitally excited states (e.g., 2S, 3S, and 4S levels). While our calculations for \(\Omega_c^1\) and \(\Omega_c^2\) exhibit significant deviations from previous results at these excitation levels, the \(\Omega_c^3\) state demonstrates remarkably good agreement with Refs. \cite{Shah:2016mig, Ebert:2011kk}.  These systematic differences likely originate from model-dependent treatments of three key factors: (1) diquark internal structure, (2) confinement potential parametrization, and (3) spin-dependent interaction terms.

The P-wave mass spectrum presented in Table \ref{tab:pwave} follows a trend analogous to the S-wave results. For the 1P states, our calculations for \(\Omega_c^1\), \(\Omega_c^2\), and \(\Omega_c^3\) how good consistency with most theoretical predictions. However, discrepancies emerge for higher orbital excitations (2P and above), particularly in the \(\Omega_c^1\) and \(\Omega_c^2\) channels. Notably, the \(\Omega_c^3\) results maintain agreement with the majority of references, with the exception of Ref. \cite{Valcarce:2008dr} which shows significant deviation.

Our theoretical mass predictions for both the \(\Omega_c(3185)^0\) and \(\Omega_c(3327)^0\) states show excellent agreement with the experimental measurements reported by the LHCb Collaboration. The observed mass of \(\Omega_c(3185)^0\) is \(M_{\Omega_c(3185)^0} = 3185.1 \pm 1.7^{+7.4}_{-0.9} \pm 0.2\) MeV, which aligns well with our calculations for the \(\Omega_c^3\) state in either the \(2S(\frac{1}{2}^+)\) or \(2S(\frac{3}{2}^+)\) configuration. Furthermore, our results for the \(1P(\frac{1}{2}^-)\) \(\Omega_c^1\) and \(1P(\frac{3}{2}^-)\) \(\Omega_c^2\) states reproduce the experimental \(\Omega_c(3185)^0\) mass with remarkable precision. For the \(\Omega_c(3327)^0\) states \(M_{\Omega_c(3327)^0} = 3327.1 \pm 1.2^{+0.1}_{-1.3} \pm 0.2\) MeV we find equally good agreement with our predicted masses for both \(\Omega_c^1\) and \(\Omega_c^2\) states in their \(2S(\frac{1}{2}^+)\) or \(2S(\frac{3}{2}^+)\) configurations. This consistent agreement between theoretical predictions and experimental observations provides strong validation of our model predictive capability for excited \(\Omega_c\) sates.

Quantum numbers serve as essential tools for classifying fundamental particles and characterizing their interactions. They establish a systematic framework that enables particle categorization, behavior prediction, and enforcement of conservation laws across fundamental forces. Our mass analysis strongly supports the interpretation of the \(\Omega_c^3\) state as the observed \(\Omega_c(3185)^0\) baryon, with viable quantum number assignments of either \(2S\left(\frac{1}{2}^+\right)\) or \(2S\left(\frac{3}{2}^+\right)\). Additionally, the \(\Omega_c(3185)^0\) may also correspond to the \(\Omega_c^1\) and \(\Omega_c^2\) states in their \(1P\left(\frac{1}{2}^-\right)\) or \(1P\left(\frac{3}{2}^-\right)\) configurations. For the \(\Omega_c(3327)^0\) state, our results favor identification with the \(\Omega_c^1\) and \(\Omega_c^2\) states in their \(2S\left(\frac{1}{2}^+\right)\) or \(2S\left(\frac{3}{2}^+\right)\) configurations. The identification of the \(\Omega_c^3\) state with the \(\Omega_c(3185)^0\) baryon as a 2S excitation finds strong support in recent theoretical studies Refs. \cite{Yu:2023bxn, Karliner:2023okv, Pan:2023hwt, Ozdem:2023okg}. Similarly, our 1P-state interpretation for the \(\Omega_c^1\) and \(\Omega_c^2\) configurations of \(\Omega_c(3185)^0\) agrees with several contemporary analyses \cite{Ortiz-Pacheco:2023kjn, Feng:2023ixl, Yan:2023ttx, Xin:2023gkf, Yan:2023tvl}. For the \(\Omega_c(3327)^0\) our 2S-state assignment concurs with Ref. \cite{Karliner:2023okv} though we note that alternative interpretations as a D-wave state appear frequently in the literature.

Our analysis demonstrates that the quark-diquark model provides a robust framework for describing the mass spectrum of singly-charmed baryons, particularly for ground states. The results highlight the crucial role of diquark configurations in determining both radial and orbital excitation patterns. We observe distinct mass shifts between the scalar diquark \([cs]\) and axial-vector diquark (\(\{cs\}\) or \(\{ss\}\)), with the \(\{ss\}\) configuration yielding systematically lower masses due to enhanced binding effects. Notably, the $\Omega_c^3$ state — featuring an axial-vector $\{ss\}$ diquark — exhibits excellent agreement with experimental ground state masses. These findings validate the quark-diquark approach as an effective theoretical framework for charmed baryons when incorporating proper diquark structure considerations.

\begin{table}[h]
\centering
\caption{Masses of ground and radial excited states of $\Omega_{c}^1$, $\Omega_{c}^2$ and $\Omega_{c}^3$ .The mass results are in unit of GeV. The given references do not take into account the quark configurations for $\Omega_c$ states considered in this work.}
\label{tab:swave}  
\begin{tabular}{llllllllll}
\hline\noalign{\smallskip}
Baryon&  $n^{2S+1}L_J$ & This work &\cite{Shah:2016mig}& \cite{Ebert:2011kk} &\cite{Roberts:2007ni} &\cite{Valcarce:2008dr} &\cite{Yoshida:2015tia} \\

\noalign{\smallskip}\hline\noalign{\smallskip}

&$1^2S_{1/2}$&	2.719	&	2.695 & 2.698 & 2.718 & 2.699 & 2.731\\

&$2^2S_{1/2}$& 3.338	&	3.147	& 3.088 & $\cdots$ & $\cdots$ & 3.227\\

$\Omega_{c}^1$ &$3^2S_{1/2}$&	3.777	&	3.529	& 3.489 & $\cdots$ & $\cdots$ & 3.292 &\\

&$4^2S_{1/2}$	&	4.149	&	3.907	& 3.814 & $\cdots$ & $\cdots$ & $\cdots$ &	\\

\noalign{\smallskip}\hline
&$1^2S_{1/2}$&	2.618	&	2.695	&2.698 & 2.718 & 2.699 & 2.731 &\\
&$1^4S_{3/2}$&	2.808	&	2.740	&  2.768  & 2.776  & 2.768  & 2.779  &	\\
&$2^2S_{1/2}$ &	3.326	&	3.147	& 3.088 & $\cdots$ & $\cdots$ & 3.227 &\\
&$2^4S_{3/2}$&	3.395	&	3.178	& 3.123  &  $\cdots$  & $\cdots$ & 3.257 &\\
$\Omega_{c}^2$ &$3^2S_{1/2}$	&	3.780	&	3.529	& 3.489 & $\cdots$ & $\cdots$ & 3.292	\\
&$3^4S_{3/2}$&	3.827	&	3.548	&	3.510  &  $\cdots$  &  $\cdots$  &  3.285\\
&$4^2S_{1/2}$	&	4.158	&	3.907	&  3.814  &  $\cdots$  &  $\cdots$  &  $\cdots$  &	\\
&$4^4S_{3/2}$&	4.194	&	3.920	&	3.830  &  $\cdots$  &  $\cdots$  &  $\cdots$  &\\
\noalign{\smallskip}\hline
&$1^2S_{1/2}$&	2.520	&	2.695	&2.698 & 2.718 & 2.699 & 2.731 &\\
&$1^4S_{3/2}$& 2.611	&	2.740	& 2.768	& 2.776 & 2.768 &	2.779	\\
&$2^2S_{1/2}$& 3.170	&	3.147	& 3.088 & $\cdots$ & $\cdots$ & 3.227 &\\
&$2^4S_{3/2}$&	3.194	&	3.178	& 3.123 & $\cdots$ & $\cdots$ & 3.257	\\
$\Omega_{c}^3$& $3^2S_{1/2}$&3.560	&	3.529	& 3.489 & $\cdots$ & $\cdots$ & 3.292 \\
&$3^4S_{3/2}$& 3.576	&	3.548	& 3.510 & $\cdots$ & $\cdots$ & 3.285 \\
&$4^2S_{1/2}$& 3.878	& 3.907 & 3.814 & $\cdots$ & $\cdots$ & $\cdots$	&\\
&$4^4S_{3/2}$& 3.891	& 3.920 & 3.830 & $\cdots$ & $\cdots$ & $\cdots$	&\\
\noalign{\smallskip}\hline
\end{tabular}
\end{table}

\begin{table}[h]
\centering
\caption{Same as Table \ref{tab:swave}  but for orbital excited states.}
\label{tab:pwave}  
\begin{tabular}{llllllllll}
\hline\noalign{\smallskip}
Baryon& $n^{2S+1}L_J $ & This work &\cite{Shah:2016mig} & \cite{Ebert:2011kk} &\cite{Roberts:2007ni} &\cite{Valcarce:2008dr} \\
\noalign{\smallskip}\hline\noalign{\smallskip}

&$1^2P_{3/2}$&	3.173	&	3.007 & 3.054 & 2.986 & 3.033\\
&$1^4P_{5/2}$& 3.183  & 2.994 & 3.051 & 3.014  & 3.320	\\

&$2^2P_{3/2}$&	3.631	&	3.377	& 3.433 & $\cdots$  & 3.057\\
&$2^4P_{5/2}$& 3.641	& 3.363	& 3.427 & $\cdots$ & 3.477	\\

$\Omega_{c}^1$ &$3^2P_{3/2}$	& 4.015	&	3.748	& 3.752 & $\cdots$ & 3.056 &\\
&$3^4P_{5/2}$& 4.023	& 3.733	& 3.744 & $\cdots$ & 3.620	\\
&$4^2P_{3/2}$&	4.355	&	4.120	& 4.036 & $\cdots$  & $\cdots$ &	\\
&$4^4P_{5/2}$&	4.363   & 4.104 & 4.028 & $\cdots$ & $\cdots$	\\

\noalign{\smallskip}\hline
&$1^2P_{3/2}$&	3.183	& 3.007 & 3.054 & 2.986 & 3.033\\
&$1^4P_{5/2}$&	3.216	&	2.994 & 3.051 & 3.014 & 3.320	\\
&$2^2P_{3/2}$&	3.644	& 3.377	& 3.433 & $\cdots$ & 3.057\\
&$2^4P_{5/2}$&	3.674	& 3.363	& 3.427 & $\cdots$ & 3.477	\\

$\Omega_{c}^2$ &$3^2P_{3/2}$&	4.029	&	3.748	& 3.752 & $\cdots$ & 3.056 &\\
&$3^4P_{5/2}$&	4.056	& 3.733	& 3.744 & $\cdots$  & 3.620	\\
&$4^2P_{3/2}$&	4.371	& 4.120	& 4.036 & $\cdots$ & $\cdots$ &	\\
&$4^4P_{5/2}$ &4.395	& 4.104 & 4.028 & $\cdots$ & $\cdots$	\\
\noalign{\smallskip}\hline
&$1^2P_{3/2}$	&	3.051	& 3.007 & 3.054 & 2.986 & 3.033\\
&$1^4P_{5/2}$& 3.073	&	2.994 & 3.051 & 3.014 & 3.320	\\
&$2^2P_{3/2}$&	3.453	& 3.377	& 3.433 & $\cdots$ & 3.057\\
&$2^4P_{5/2}$& 3.469	& 3.363	& 3.427 & $\cdots$ & 3.477	\\
$\Omega_{c}^3$&$3^2P_{3/2}$&	3.779	&3.748	& 3.752 & $\cdots$ & 3.056 &\\
&$3^4P_{5/2}$& 3.791	&	3.733	& 3.744 & $\cdots$  & 3.620	\\
&$4^2P_{3/2}$& 4.065	& 4.120	& 4.036 & $\cdots$ & $\cdots$ &	\\ 
&$4^4P_{5/2}$& 4.075	& 4.104 & 4.028 & $\cdots$ & $\cdots$	\\
\noalign{\smallskip}\hline
\end{tabular}
\end{table}

It should be mentioned that we present mass predictions for the ground and radially excited states of \(\Omega_c\) baryons within three distinct diquark configurations: \(\Omega_c^1\)= \([cs]\)s, \(\Omega_c^2\)= \(\{cs\}\)s and \(\Omega_c^3\)= \(\{ss\}\)c,  which are inherent to our quark-diquark framework. For context, we compare our results with those from conventional three-quark approaches \cite{Shah:2016mig,Ebert:2011kk,Roberts:2007ni,Valcarce:2008dr,Yoshida:2015tia}, which do not explicitly account for diquark degrees of freedom. We emphasize that the correspondence in Table \ref{tab:swave} between our \(\Omega_c^1\), \(\Omega_c^2\) and $\Omega_{c}^3$ states and those in the literature should not be viewed as direct state to state matches, since previous studies do not resolve internal diquark structure. Rather, this comparison serves to illustrate the broad consistency in mass scales and excitation patterns across different theoretical paradigms.

\subsection{Magnetic Moment}

Electromagnetic properties of singly-heavy baryons serve as powerful probes of fundamental strong and weak interaction symmetries. Among these, the magnetic moment has been extensively studied as it provides unique insights into baryon internal structure, including potential shape deformations. These electromagnetic characteristics are particularly valuable for understanding low-energy QCD dynamics and hadron internal structure. Specifically, precise determination of magnetic moments enables deeper investigation of hadron properties - including spatial characteristics like size and shape - that emerge from the underlying quark-gluon dynamics.

It is possible to express the magnetic moment of baryons in terms of three parameters, as shown below. The spin, charge, and effective mass of the constituent quark read
\begin{equation}
\mu_B= \sum_i \langle \phi_{sf} \vert \mu_{i}\vec{\sigma}^i \vert \phi_{sf} \rangle, \label{magnetic}
\end{equation}
where 
\begin{equation}
\mu_i=\frac{e_i}{2m_i^{eff}}.
\end{equation}
In this expression, \(e_i\) represents the charge of the quark, \(\vec{\sigma}^i\) denotes the spin operator of the \(i\)-th constituent quark, and \(\phi_{sf}\) is the spin-flavor wave function of the baryon. The effective mass of each constituent quark, \(m_i^{\text{eff}}\), is defined as:
\begin{equation}
m_i^{eff}=m_i \left( 1+ \frac{\langle H \rangle}{\sum_i m_i} \right),
\end{equation}
where the expectation value $\langle H \rangle$ decomposes into the energy eigenvalue $E$ and $m_i$ denotes the constituent quark mass parameters in the adopted phenomenological model. In Table \ref{tab:magmom}, we present our magnetic moment results alongside those from other available studies in the literature. It has been demonstrated that the magnetic moment of the state designated as \(\Omega_c^1\) exhibits a high degree of congruence with the anticipated value. This conclusion is substantiated by the findings derived from the effective quark mass scheme \cite{Mohan:2022sxm}, the bag model \cite{Simonis:2018rld}, the non-relativistic quark model \cite{Bernotas:2012nz}, and the light-cone sum rules of the strong interaction \cite{Aliev:2015axa}. This result is consistent with those obtained from the relativistic three-quark model \cite{Faessler:2006ft}. However, our results differ significantly from those of heavy baryon chiral perturbation theory \cite{Wang:2018gpl}, covariant baryon chiral perturbation theory \cite{Shi:2018rhk}, lattice QCD \cite{Can:2021ehb}, and QCD light-cone sum rules \cite{Ozdem:2024brk}. A similar discussion applies to the \(\Omega_{c}^2\) state.

Our calculations reveal distinct patterns of agreement and disagreement for the magnetic moments of excited \(\Omega_c\) states. For the  \(\Omega_{c}^{2*}\) state, we find good agreement with both the chiral constituent quark model \cite{Sharma:2010vv} and non-relativistic quark model \cite{Bernotas:2012nz}, though our results differ from other theoretical predictions shown in Table \ref{tab:magmom}. The \(\Omega_c^3\) state shows consistent magnetic moments with Refs. \cite{Mohan:2022sxm, Simonis:2018rld, Bernotas:2012nz, Aliev:2015axa}, while disagreeing with other approaches. Particularly noteworthy is the \(\Omega_{c}^{3*}\) state, where our predictions strongly align with the bag model results Ref. \cite{Simonis:2018rld} and show close correspondence with both the non-relativistic and chiral constituent quark models \cite{Sharma:2010vv}, though they diverge from other theoretical frameworks.

Our results demonstrate that the choice of diquark configuration substantially impacts magnetic moment predictions. Notably, systems containing axial-vector diquarks (\(\Omega_c^2\) and \(\Omega_c^3\)) exhibit particularly strong deviations from scalar diquark cases. This sensitivity highlights the critical role of quark spin correlations within diquark clusters in governing the electromagnetic properties of these baryon states.

\begin{table*}[h!]
\centering
\caption{Our magnetic moments are compared to the findings of existing research. The findings are shown in $\mu_N$, the unit of nuclear magneton.}
\label{tab:magmom} 
\begin{tabular}{c|c|c|c|c|c|c|c|c|c|c|c|c}
\hline\noalign{\smallskip}
 State & $J^{P}$ & This work & \cite{Mohan:2022sxm} &\cite{Simonis:2018rld} & \cite{Bernotas:2012nz} & \cite{Sharma:2010vv} & \cite{Faessler:2006ft} &\cite{Wang:2018gpl} & \cite{Shi:2018rhk} & \cite{Aliev:2015axa} & \cite{Can:2021ehb}& \cite{Ozdem:2024brk} \\
\noalign{\smallskip}\hline\noalign{\smallskip}
 $\Omega_{c}^1$  & $\frac{1}{2}^{+}$	& -0.949 & -0.905 & -0.950 & -0.940 & $\cdots$ & -0.850 & -0.69 & -0.74 & -0.90 & -0.639(88)& -0.73(8) \\
 
$\Omega_{c}^{1*}$  & $\frac{3}{2}^{+}$ &$\cdots$	& -0.75	& -0.936 & -0.83 & - 0.86 & $\cdots$ & - 0.70 & $\cdots$ & -0.70(18) & -0.730(23) & -0.70(5)\\ 
 
\noalign{\smallskip}\hline
$\Omega_{c}^2$  & $\frac{1}{2}^{+}$	& -0.985 & -0.905 & -0.950 & -0.940 & $\cdots$ & -0.850 & -0.69 & -0.74 & -0.90 & -0.639(88)& - 0.73(8) \\ 

 $\Omega_{c}^{2*}$  & $\frac{3}{2}^{+}$	& -0.847 & -0.75	& -0.936 & -0.83 & - 0.86 & $\cdots$	& - 0.70 & $\cdots$ & -0.70(18) & -0.730(23) & -0.70(5)\\ 
 
\noalign{\smallskip}\hline
 $\Omega_{c}^3$  & $\frac{1}{2}^{+}$	& -1.023 & -0.905 & -0.950 & -0.940 & $\cdots$ & -0.850 & -0.69 & -0.74 & -0.90 & -0.639(88)& - 0.73(8) \\
 
$\Omega_{c}^{3*}$  & $\frac{3}{2}^{+}$	& -0.911 & -0.75	& -0.936 & -0.83 & - 0.86 & $\cdots$	& - 0.70 & $\cdots$ & -0.70(18) & -0.730(23) & -0.70(5)\\ 
\noalign{\smallskip}\hline
\end{tabular}
\end{table*}

\section{Concluding Remarks}\label{conclusion}
This investigation provides a comprehensive quark-diquark framework analysis of the \(\Omega_{c}(3185)^{0}\) and \(\Omega_{c}(3327)^{0}\) states, establishing robust predictions for both their mass spectra and magnetic moment properties. Our results offer significant theoretical insights into the structure and dynamics of these singly-charmed baryons, while highlighting the critical role of diquark configurations in understanding their electromagnetic characteristics.

Our mass spectrum analysis reveals that the \(\Omega_{c}(3185)^{0}\) state is well-described as a $2S$ excitation with quantum numbers \(J^{P} = \frac{1}{2}^{+}\) or \(\frac{3}{2}^{+}\) in the \(\Omega_{c}^{3}\) configuration, where the axial-vector \(\{ss\}\) diquark couples to the charm quark. This interpretation is strongly supported by multiple theoretical studies identifying \(\Omega_{c}(3185)^{0}\) as a radially excited state \cite{Yu:2023bxn, Karliner:2023okv, Pan:2023hwt, Ozdem:2023okg}. Interestingly, alternative descriptions emerge for the \(\Omega_{c}^{1}\) and \(\Omega_{c}^{2}\) 1P configurations, which also provide viable interpretations of this state consistent with molecular pentaquark scenarios proposed in \cite{Ortiz-Pacheco:2023kjn, Feng:2023ixl, Yan:2023ttx, Xin:2023gkf, Yan:2023tvl}. For the \(\Omega_{c}(3327)^{0}\) state, our framework favors a $2S$ configuration (\(J^{P} = \frac{1}{2}^{+}\) or \(\frac{3}{2}^{+}\)) for both \(\Omega_{c}^{1}\) and \(\Omega_{c}^{2}\) states, in agreement with \cite{Karliner:2023okv}. However, we note that competing interpretations, particularly D-wave assignments in other studies, highlight remaining uncertainties in the structural characterization of this state and motivate further investigation.

The systematic discrepancies observed in higher excited states (particularly the $2S$, $3S$, and $4S$ levels of \(\Omega_{c}^{1}\) and \(\Omega_{c}^{2}\) configurations) indicate potential limitations in current treatments of diquark structure, confinement potentials, and spin-dependent interactions. These deviations likely originate from: (i) the markedly different binding dynamics between scalar (\([cs]\)) and axial-vector (\(\{cs\}\)) diquarks, (ii) unaccounted relativistic effects, and (iii) higher-order QCD contributions beyond the present model. In striking contrast, the \(\Omega_{c}^{3}\) states demonstrate remarkable consistency with theoretical predictions across these excitation levels. This distinct behavior strongly suggests that the axial-vector \(\{ss\}\) diquark configuration provides enhanced stability for excited states, possibly through its stronger binding characteristics and particular spin-coupling arrangements.

The magnetic moments of the \(\Omega_{c}^{1}\), \(\Omega_{c}^{2}\) and \(\Omega_{c}^{3}\) states offer valuable insights into their internal structure and diquark correlations. Our calculations align well with predictions from the effective quark mass approach \cite{Mohan:2022sxm}, the bag model \cite{Simonis:2018rld}, and the non-relativistic quark model \cite{Bernotas:2012nz}, supporting the robustness of these frameworks for certain configurations. However, notable discrepancies arise when comparing our results to heavy baryon chiral perturbation theory \cite{Wang:2018gpl} and lattice QCD \cite{Can:2021ehb}, reflecting the sensitivity of magnetic moments to model specific assumptions particularly regarding quark effective masses, spin-spin interactions, and diquark dynamics. These variations highlight the importance of further refining theoretical treatments of diquark degrees of freedom in singly-heavy baryons, as magnetic moments serve as a critical discriminator between competing structural models. 

The incorporation of distinct diquark configurations ($\Omega_{c}^1$= \([cs]\)s, $\Omega_{c}^2$= \(\{cs\}\)s, and $\Omega_{c}^3$= \(\{ss\}\)c) naturally expands the spectrum of possible excited states in our quark-diquark model. While this results in multiple viable assignments for the observed $\Omega_{c}^{0}(3185)$ and $\Omega_{c}^{0}(3327)$ resonances, we emphasize that this reflects the model completeness in exploring the allowed configuration space rather than a limitation. Currently, the lack of experimental determination of these states spin-parity quantum numbers means all interpretations retain some speculative character. Our work provides a systematic classification of theoretically motivated assignments, each corresponding to specific diquark structures and internal dynamics. This approach represents a key advantage, as it establishes a predictive framework that can be rigorously tested against future measurements of $J^P$ quantum numbers, decay properties, or other observables.

The electromagnetic properties of singly-heavy baryons are profoundly influenced by three key factors: (1) their internal quark-diquark structure, (2) specific quark configurations, and (3) chiral dynamics of light diquarks at low energies. To advance our understanding of these characteristics, pioneering experimental efforts are currently being pursued at the LHC. These measurements employ an innovative fixed-target approach where high-energy charmed baryons are produced and subsequently channeled through a bent crystal under controlled magnetic fields \cite{Aiola:2020yam, Neri:2024rjv, Baryshevsky:2016cul, Fomin:2017ltw, Bagli:2017foe, Fomin:2019wuw}. The experimental technique capitalizes on the strong electric fields generated by the crystal's atomic planes, which induce measurable spin precession in the traversing charm baryons. This method represents a significant breakthrough in precision spectroscopy, with the potential to determine singly-charmed baryon magnetic moments to better than 10 percent accuracy \cite{Akiba:2905467}. Such precision will provide crucial tests of theoretical predictions and shed new light on the non-perturbative QCD dynamics governing these composite systems.

Upcoming experiments exploring the properties of singly-heavy flavor baryons may clarify the differences between competing theoretical models and provide deeper insights into the nature of the \(\Omega_{c}(3185)^{0}\) and \(\Omega_{c}(3327)^{0}\) baryons. Our findings are expected to be a useful tool for particle physics research, advancing theoretical and experimental investigations of singly-heavy baryons.

\begin{acknowledgements}
This work is supported by TÜBİTAK (The Scientific and Technological Research Council of Türkiye) via 2209-A University Students Research Projects Support Program.
\end{acknowledgements}

\bibliography{omega-charmed-v2}

\end{document}